# Analytical evaluation of the lattice Green function for the face centered cubic lattice

B.A.Mamedov

*Department of Physics, Faculty of Arts and Sciences, Gaziosmanpaşa University, Tokat, Turkey*

**Abstract**

An efficient calculation method is proposed for the face centered cubic (FCC) lattice Green function. The method is based on binomial expansion theorems, which is provide us establish analytical formulae through simple basic integrals. The resulting series present better convergence rates. Several acceleration techniques are combined to further improve the efficiency. The obtained results for the lattice Green function are in good agreement with the known numerical calculation results.

**Key words:** Lattice Green functions; face centered cubic lattice; anisotropic interaction; binomial coefficients

## I. Introduction

The lattice Green function play a key role in the theory of solid state physics [1, 2]. An understanding of the structure of lattice Green function is frequently encountered in lattice statistical problems [2-5]. In literature, many efficient approaches have been reported for these functions [6-9]. As seen from literature that most of the studies on lattice functions are based on elliptic integral and recurrence relation approaches [6-9]. Unfortunately, for most of these purely elliptic integral and recurrence relation methods, there are some limitations in their applicability despite the huge increase in the computational resources. We have presented a detailed analytical formulae for the computation FCC lattice Green function by using binomial expansion theorems. The obtained simple analytical formula for the lattice Green function is completely general for $t \geq 3$.

In this paper, a new approach to the computation of the lattice Green functions are proposed, which considerable improved its capabilities during numerical evaluation in significant cases. The new analytical approach method for evaluating the lattice Green function is conceptually simpler than existing methods in the literature.

## 2. Definition and basic formulas

The lattice Green function for fcc lattice is defined as

$$G(t,l,m,n) = \frac{1}{\pi^3} \int_0^\pi \int_0^\pi \int_0^\pi \frac{\cos lx \cos my \cos ny}{t - \omega(x,y,z)} dxdydz \qquad (1)$$

$$\omega(x,y,z) = \gamma \cos x \cos y + \cos y \cos z + \cos z \cos x \qquad (2)$$

where $t$ is a complex number, which is described in terms of energy in solid state physics, and $(l,m,n)$ is such a set of integer that the sum $l+m+n$ is an even number [8]. $\gamma$ is the parameter which is unity for the isotropic FCC lattice.

In order to establish expressions for the lattice Green function fist we shall consider well known binomial expansion theorems as follows [10-12]:

$$(x \pm y)^n = \lim_{N' \to \infty} \sum_{m=0}^{N'} (\pm 1)^m F_m(n) x^{n-m} y^m, \qquad (3)$$

Here $N$ is the upper limits of summations and $F_m(n)$ are binomial coefficients defined by

$$F_m(n) = \begin{cases} \dfrac{n(n-1)...(n-m+1)}{m!} & \text{for integer } n \\ \dfrac{(-1)^m \Gamma(m-n)}{m! \Gamma(-n)} & \text{for noninteger } n \end{cases} \qquad (4)$$

We notice that for $m < 0$ the binomial coefficient $F_m(n)$ in Eq. (3) is zero and the positive integer $n$ terms with negative factorials do not contribute to the summation. Taking into account Eq.(3) in Eq.(1) we obtain for the lattice Green function the series expansion formulas in terms of binomial coefficients and basic integrals:

$$G(t,l,m,n) = \frac{1}{\pi^3} \lim_{N \to \infty} \sum_{i=0}^{N} (-1)^i F_i(-1) t^{-1-i} \sum_{j=0}^{i} F_j(i) \gamma^{i-j} \sum_{k=0}^{j} F_k(j) J_{i-j+k}(l) J_{i-k}(m) J_j(n) \text{ for } t \geq 3, (5)$$

We can evaluate the lattice Green function using alternative formulae:

$$G(t,l,m,n) = \frac{1}{\pi^3} \lim_{\substack{N \to \infty \\ L \to \infty}} \sum_{i=0}^{N} (-1)^i F_i(-1) t^{-1-i} \sum_{j=0}^{L} (-1)^j F_j(-1-i) \gamma^j t^{-j}$$
$$\times \sum_{k=0}^{i} F_k(i) J_{j+k}(l) J_{j+i-k}(m) J_i(n) \qquad \text{for } t \geq 3 \qquad (6)$$

where

$$J_n(k) = \begin{cases} I_n & \text{for} \quad k = 0 \\ L_n(k) & \text{for} \quad k > 1 \\ I_{n+1} & \text{for} \quad k = 1 \\ 0 & \text{for} \quad k > n \text{ or } k+n \text{ odd} \end{cases} \qquad (7)$$

The basic integrals $L_n(k)$ and $I_n$ occurring in Eq.(7) is determined from the following relations, respectively

$$L_n(k) = \int_0^\pi \cos kx \cos^n x\, dx = 2^{k-1} I_{k+n} + k \sum_{i=1}^{E[k/2]} \frac{(-1)^i 2^{k-2i-1} F_{i-1}(k-i-1) I_{k+n-2i}}{i} \qquad (8)$$

and

$$I_n = \int_0^\pi \cos^n \varphi\, d\varphi = \begin{cases} 0, & \text{if } n \text{ odd} \\ \sqrt{\pi}\, \dfrac{\Gamma\left(\dfrac{n+1}{2}\right)}{\Gamma\left(\dfrac{n}{2}+1\right)}, & \text{if } n \text{ even} \end{cases}. \qquad (9)$$

In Eq.(7) the index $E[k/2]$ is the upper limit of summation defined by

$$E(n/2) = \frac{n}{2} - \frac{1}{4}[1-(-1)^n]. \qquad (10)$$

In the present work, we propose an alternative method for the analytical evaluation of the FCC lattice Green function. The obtained formulas are practically simple and they offer some advantages over currently available methods.